\documentclass[letterpaper,12pt]{article}   
\usepackage{osajnl2} 
\usepackage[draft]{hyperref} 

\begin{document}

\title{Disturbance of Soliton Pulse Propagation from Higher-Order Dispersive Waveguides }

\author{Matthew Marko$^{1,2,*}$, Andrzej Veitia$^{2,3}$, Xiujian Li$^{2,4}$, Jiangjun Zheng$^{2}$, and Chee-Wei Wong$^{2}$}
\address{$^{1}$Navy Air Warfare Center Aircraft Division (NAWCAD), Joint Base McGuire-Dix-Lakehurst, Lakehurst NJ 08733, USA}
\address{$^2$Department of Mechanical Engineering, Columbia University in the City of New York, New York NY 10027, USA}
\address{$^3$Department of Electrical Engineering, University of California Riverside, Riverside, CA 92521, USA}
\address{$^4$Science College, National University of Defense Technology, Changsha, Hunan 410073, PRC}
\address{$^*$Corresponding author: matthew.marko@navy.mil}

\begin{abstract}Optical soliton pulses offer many applications within optical communication systems, but by definition a soliton is only subjected to second-order anomalous group-velocity-dispersion; an understanding of higher-order dispersion is necessary for practical implementation of soliton pulses.  A numerical model of a waveguide was developed using the Nonlinear Schr\"{o}dinger Equation, with parameters set to ensure the input pulse energy would be equal to the fundamental soliton energy.  Higher-order group-velocity-dispersion was gradually increased, for various temporal widths and waveguide dispersions.  A minimum pulse duration of 100-fs was determined to be necessary for fundamental soliton pulse propagation in practical photonic crystal waveguides.  
\end{abstract}

\maketitle

\section{\label{Intro} Introduction}
\indent The purpose of this effort is to, both theoretically and with numerical simulations, study the effects of higher-order dispersion on fundamental soliton pulse propagation of ultrafast pulses propagating within an optical waveguide.  Optical solitons are undistorted pulse envelopes that can propagate over long distances undistorted \cite{1,2}, which result as part of the interplay between the nonlinear Kerr effect and anomalous dispersion \cite{1,2,3,4,5,6,7}.  These pulses are of interest to the scientific community for their self-sustaining properties, which are useful both for low-powered optical data transfer as well as high-energy pulsed laser systems.  These pulses have been generated in optical fibers \cite{1,6}, where the stable nature of the pulse reduces loss and cross-talk.  This effort is to study the feasibility of self-sustaining soliton pulses in the presence of higher-order group-velocity dispersion (GVD) \cite{1}, with emphasis placed on practical studies of photonic crystals waveguides (PhCWG) \cite{3,8,9,10,11}, which are subjected to much higher-order dispersion as a result of the geometric design of the different waveguide dielectrics.  

\indent The optimal way of making such a determination, other than through experimentation, is to use numerical methods.  These methods include Finite Difference Time Domain (FDTD) and the split-step Nonlinear Schr\"{o}dinger Equation (NLSE) method \cite{1,12,13,14,15}.  The NLSE is one of the most effective methods for numerically realizing soliton pulse evolution in a nonlinear dispersive waveguide, as it is easy to implement and computationally efficient.  Just like FDTD, the NLSE method is derived from Maxwell's equation \cite{2}, under the assumption that the Slowly Varying Envelope Approximation (SVEA) is applicable \cite{2,16}, where it is assumed that the envelope of the pulse varies slowly in time and space compared to an optical period.  

\indent Soliton pulses within fiber optics is a well-established subject, and it has been used previously for data communication over many kilometers of fiber networks \cite{1,6}.  The primary disadvantage of fibers is simply the long-length scales required for the optical nonlinearities to occur.  There would be many advantages to having soliton pulse compression occur at the millimeter length scales, which would allow for better optical data transfer and processing within a typical silicon computer chip \cite{3,4,5,12,15,17,18,19,20,21,22}.  For this reason, there has been great interest in the development of photonic crystals, which possess extraordinary nonlinear characteristics at much smaller length scales.  

\indent One of the first examples of soliton pulse compression at the millimeter length scale has been observed in gallium indium phosphate (GaInP) PhCWG \cite{3}.  Ultrafast pulses measured at 2 picoseconds (ps) were propagated through these PhCWG, which are engineered to have extremely high anomalous dispersion \cite{7}, and the output pulse was measured with autocorrelation at 500 femtoseconds (fs).  These PhCWG, by their definition, were measured to have orders of magnitude greater $\mathrm{2^{nd}}$ and $\mathrm{3^{rd}}$-order GVD coefficients when compared to silica fibers \cite{3,9}.  

\indent One issue inherently manifested in photonic crystals is the robustness of NLSE and the soliton wave; due to the highly dispersive nature of photonic crystals the SVEA tends to break down quickly.  The robustness of the NLSE for few-optical cycle pulses has been studied both experimentally and numerically in optical fibers \cite{16,23}, and analytically in photonic crystals \cite{10}.  It is clear that proper engineering of the higher-order dispersion in the PhCWG is necessary for sustaining the shape of the pulse as it propagates through the waveguide.  To the authors' knowledge, however, there has been no formal study to observe when an ultrafast pulse is too short for the soliton wave to sustain itself in a highly dispersive waveguide.  For this reason, this study was conducted, where a fundamental soliton was propagated through a highly dispersive waveguide, and the sustainability was studied for various temporal durations and higher-order GVD.  

\section{Basic Theory}
In order to accurately simulate the optical soliton propagation, two crucial nonlinear characteristics are necessary; the GVD and the nonlinear Kerr effect \cite{1,2}.  The GVD coefficients, defined as $\beta$, are determined by knowing the derivatives of the refractive index over the angular frequency.  These terms are found using the following equations \cite{1}: 
\begin{eqnarray}
\beta(\omega)&=&n(\omega)\frac{\omega}{c_{0}},\\
\beta_{M}(\omega)&=&\frac{d^{M}\beta(\omega)}{d \omega^{M}},\\
\beta_{1}(\omega)&=&\frac{1}{v_{g}}=\frac{n_{g}}{c_{0}}=\frac{1}{c_{0}}(n(\omega)+\omega \frac{dn(\omega)}{d\omega}),\\
\beta_{3}(\omega)&=&\frac{d{\beta_{2}(\omega)}}{d\omega}= \frac{d^{2}{\beta_{1}(\omega)}}{d\omega^{2}}= \frac{d^{3}{\beta(\omega)}}{d\omega^{3}},\\ \nonumber
\end{eqnarray}
where $n(\omega)$ is the refractive index as a function of the angular frequency ($\omega$), $c_{0}$ (m/s) is the speed of light, $v_{g}$ (m/s) is the group velocity, and $n_{g}$ is the group index.  These terms can be determined either experimentally \cite{8,25}, approximated with the Sellmeier equation \cite{1,2}, or engineered to different values as is the case with photonic crystals \cite{9}.  \\
\indent The other component of generating soliton pulse compression is the nonlinear Kerr effect.  This results in a change of refractive index proportional to the square of the electric field, or proportional to the electric field intensity.  The change in refractive index is determined by \cite{1,2,17}:
\begin{eqnarray}
\Delta{n}=n_{2}I_{1}=\frac{3 \eta_0 \chi^{(3)}}{\epsilon_{0}n} I_{1},
\end{eqnarray}
where $n_{2}$ ($\mathrm{m^{2}}$/W) is the nonlinear Kerr coefficient, $I_{1}$ (W/$\mathrm{m^{2}}$) is the intensity of the pulse, $\emph{n}$ is the refractive index, $\epsilon_0$ is the electric permittivity in a vacuum (8.854 $\cdot$ 10$\mathrm{^{-12}}$ F/m), $\eta_0$ is the vacuum impedance ($\approx{120{\pi}}$ ${\Omega}$), and $\chi^{(3)}$ (m$\mathrm{^{2}}$/V$\mathrm{^{2}}$) is the $\mathrm{3^{rd}}$-order nonlinear susceptibility coefficient \cite{2}.  These two nonlinear parameters can combine together to form the NLSE:   \\
\begin{eqnarray}
\frac{\partial A}{{\partial z}} + {\frac{1}{v_{g}}{\frac{\partial A}{{\partial t}}}} + {\frac{j}{2}{\beta_{2}}{\frac{\partial^{2} A}{{\partial t^{2}}}}}+{\frac{j}{6}{\beta_{3}}{\frac{\partial^{3} A}{{\partial t^{3}}}}}={j{\gamma{|A|^{2}}A{\cdot}{exp(-\alpha z)}}},\\
\gamma={\frac{2{\pi}{n_{2}}}{A_{e}{\cdot}{\lambda_0}}}{\cdot}{(\frac{n_{g}}{n})^{2}},
\end{eqnarray}
where $v_g$ (m/s) is the group velocity, $\beta_{2}$ ($\mathrm{s^2/m}$) is the $\mathrm{2^{nd}}$-order GVD coefficient, $\beta_{3}$ ($\mathrm{s^3/m}$) is the $\mathrm{3^{rd}}$-order GVD coefficient, $A_{e}$ $\mathrm{(m^2)}$ is the effective area of the waveguide, $\lambda_0$ (m) is the wavelength, $\gamma$ $\mathrm{(m{\cdot}W)^{-1}}$ is the nonlinear parameter, $\alpha$ ($\mathrm{m^{-1}}$) is the linear loss coefficient (disregarded throughout this theoretical study), and \emph{j}=$\sqrt{-1}$.  \\
\indent When the pulse is focused by its own intensity, it is known as Self-phase Modulation (SPM), and when there is strong SPM interplaying with strong anomalous GVD, then soliton pulse propagation may occur \cite{3}.  The SPM will cause the leading edge (longer wavelengths) frequencies to be lowered, and the trailing edge (shorter wavelengths) frequencies to be raised.  At the same time, the anomalous GVD will cause the lowered leading edge frequencies to slow down, and the raised trailing edge frequencies to speed up.  This results in the pulse narrowing, or compressing into what is known as the soliton pulse \cite{1,2,3,19}.  

\section{Numerical Modelling}
The NLSE method is also known as the split-step method, as each incremental distance step involves two fundamental steps.  The first step is to account for the GVD.  The code will, using the Fast Fourier Transforms (FFT) technique, convert the pulse information into the spectral domain, apply the GVD, and then convert the pulse back from the spectral domain to the temporal domain.  The equation for the GVD is as follows \cite{1,2,14,26,27}:
\begin{eqnarray}
\bar{A}(z,\omega)&=&\bar{A}(0,\omega){\cdot}{exp}[i \frac{1}{2}\beta_{2}\omega^{2} z+ i \frac{1}{6}\beta_{3}\omega^{3}z+...],\\
{\bar{A}}(z,\omega) &=& \int_{-\infty}^\infty {{A}(z,T)}{\cdot}{{exp}[-2{\pi}{i}{\omega}{T}]}\,\mathrm{d}T, \\
T &=& t - {\frac{z}{v_g}}, \\
\nonumber
\end{eqnarray}
where $\emph{A}(z,T)$ and $\bar{A}(z,\omega)$ are the pulse envelope functions in the temporal and spectral domains, $\emph{T}$ (s) is the normalized time, $\emph{z}$ (m) is the propagation distance, and $\omega$ (rad/s) is the normalized angular frequency.  The second step is to convert this dispersed pulse, convert it to the temporal domain, and apply the effects of SPM.  The equation for the nonlinear phase shift is \cite{1,2,14}: 
\begin{equation}
  A(z,T)=A(0,T){\cdot}{exp}[i{(\frac{2\pi}{\lambda_{0}})}{\cdot}{n_{2}}{\cdot}{{|A(0,T)|}^{2}} ].
   \end{equation}
By following through these two steps with each distance increment, one can predict the nonlinear effects on a soliton pulse envelope during and after optical propagation in a waveguide.  \\

\indent The first step in this effort was to develop a numerical model to solve the NLSE.  What was unique about the model used in this study was that it adjusted the input pulse energy to consistently propagate a fundamental soliton, depending on the related parameters such as the $\mathrm{2^{nd}}$-order GVD, the Kerr coefficient, effective area, wavelength, group index, effective index, and input temporal duration.  A fundamental soliton is a pulse with just the right energy in order that the nonlinear Kerr effect perfectly balances with the anomalous GVD, and therefore the pulse amplitude does not change as it propagates.  The input power was calculated by the equation \cite{1,3}: \\
\begin{equation}
\label{eq:MyEquToCite2}
  E_{P}=\frac{2K|\beta_{2}|}{|\gamma|{\cdot}{\tau}},\
   \end{equation}
where $\emph{K}$ is a constant to account for the numerical error, and $\tau$ (s) is the pulse temporal duration.  As expected, it was observed that for a given soliton number, $N^{2}$ = Pulse Energy / Fundamental Soliton Energy \cite{3}, the pulse shape would look identical after a given number of nonlinear lengths.  
\begin{equation}
\label{eq:MyEquToCite3}
  L_{NL} = L / \frac{{\tau}^{2}}{|\beta_{2}|},\
   \end{equation}
where $\emph{L}$ (m) is the waveguide length.

\begin{figure}[h]
\centering
\begin{minipage}[b]{0.30\linewidth}
\centering
\includegraphics[width=\textwidth]{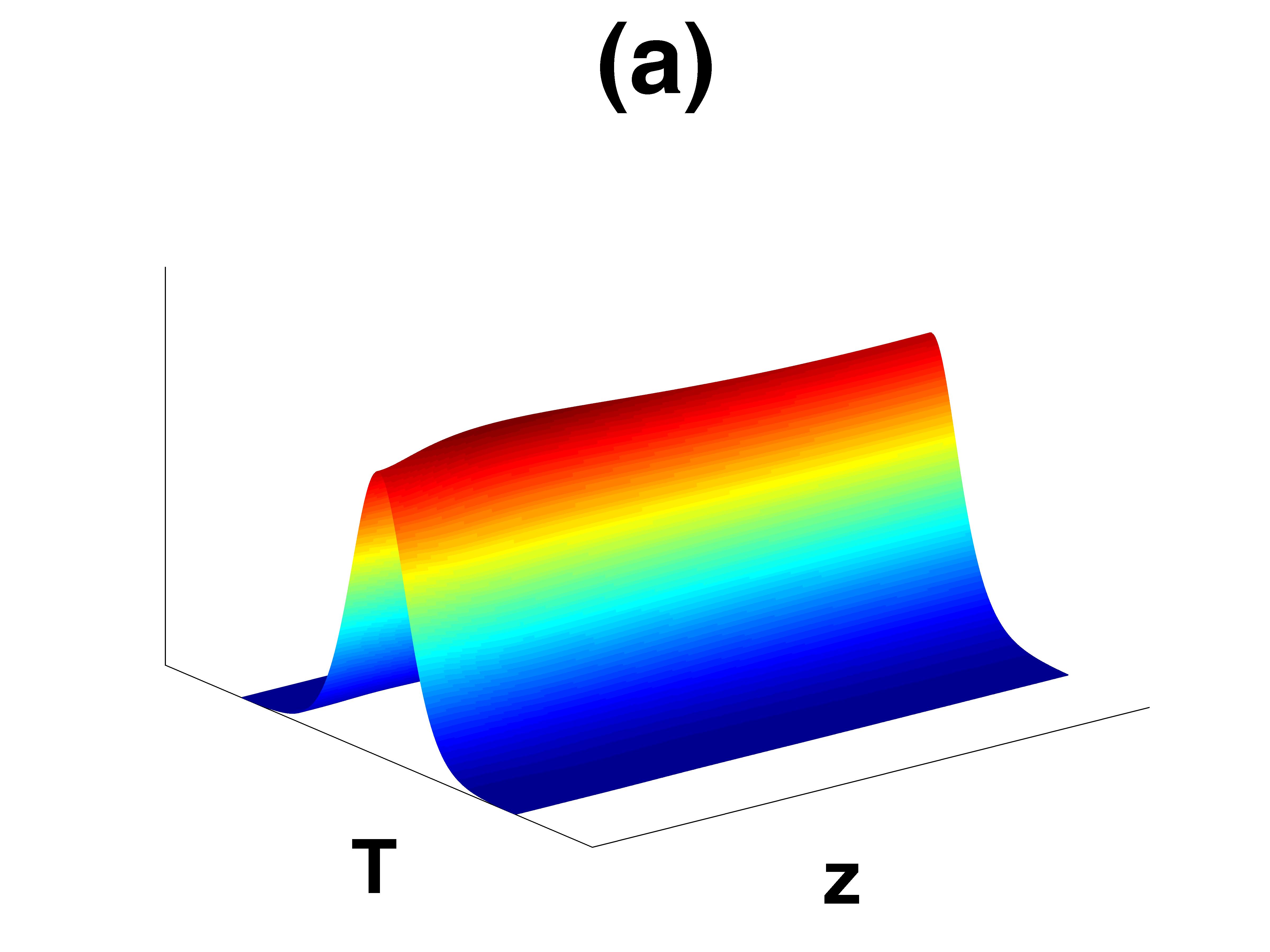}
\end{minipage}
\begin{minipage}[b]{0.30\linewidth}
\centering
\includegraphics[width=\textwidth]{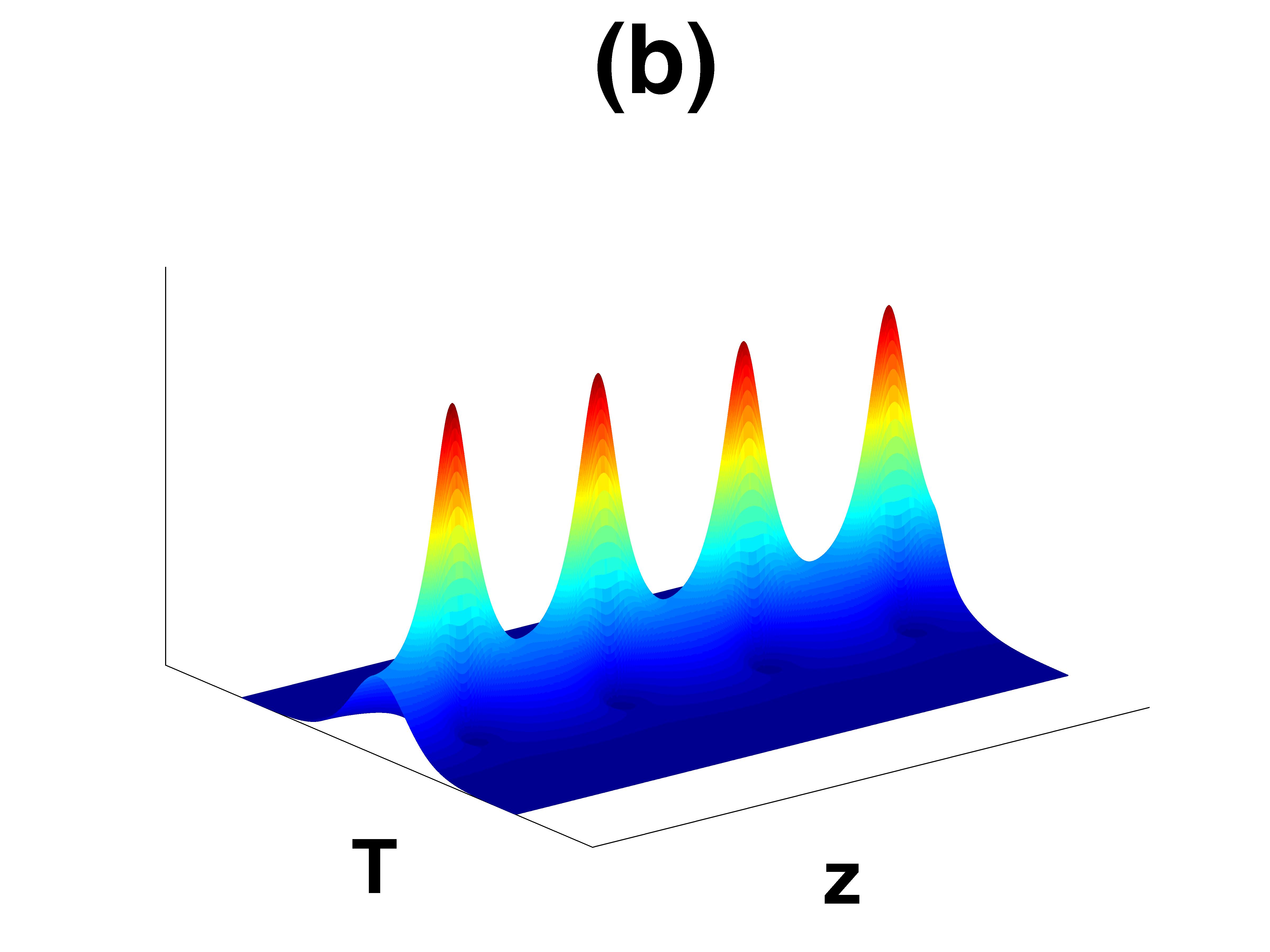}
\end{minipage}
\begin{minipage}[b]{0.30\linewidth}
\centering
\includegraphics[width=\textwidth]{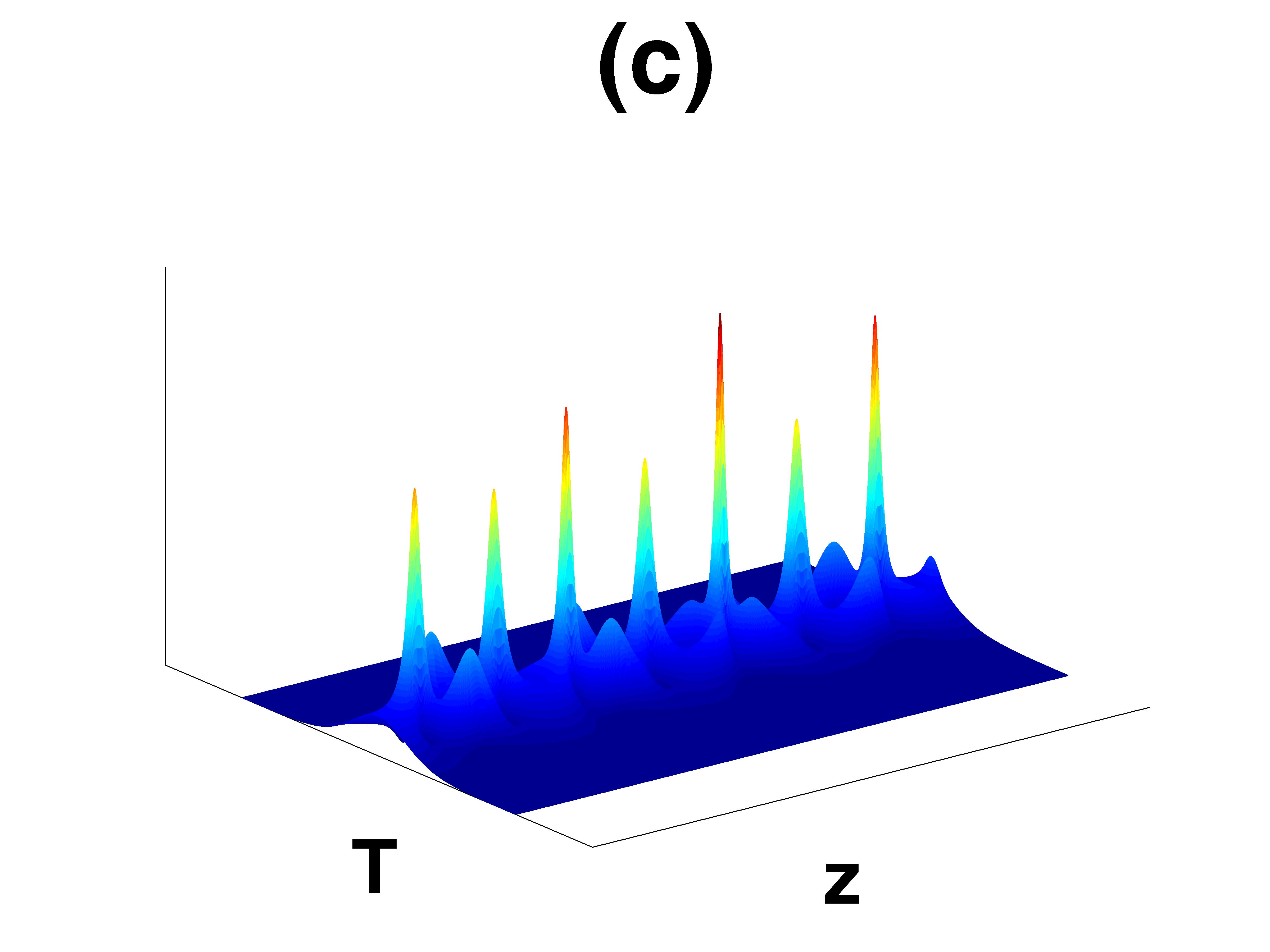}
\end{minipage}
\caption{Ideal Soliton Propagation, for the (a) fundamental soliton N = 1, (b) higher order soliton N = 2, and (c) higher order soliton N = 3.  }
\label{fig:fig1}
\end{figure}

\indent As expected, the self-focusing from the Kerr nonlinearity focused the pulse, and canceled out the anomalous dispersion of the waveguide.  In the case of the fundamental soliton, the pulse experienced little change over distance, whereas with second and third order solitons there was compression followed by splitting, with the pulse repeating this phenomena while propagating through the waveguide.  The results were verified to be the result of soliton propagation, as the pulse broadened dramatically when the Kerr coefficient was turned off as the pulse propagated in the waveguide.  \\
\indent An interesting property of the fundamental soliton that was observed numerically was the fact that, for a specified soliton number and nonlinear length, changing any of the important nonlinear parameters, such as the $\mathrm{2^{nd}}$-order dispersion, the Kerr coefficient, effective area, wavelength, group index, effective index, or temporal duration, the pulse would experience no change in shape; only the scales would change.  This is assuming there are no losses, whether linear or nonlinear such as two-photon and free-carrier absorption \cite{3,15}, as well as no higher order dispersions.  While the effects of $\mathrm{2^{nd}}$-order GVD will inherently be greater with shorter pulses (which mathematical must have larger spectral widths \cite{28}), by adjusting the energy accordingly to maintain the fundamental soliton, the pulse may continue to be self-sustaining.  The model demonstrated, based on the theory of NLSE, that a soliton can propagate through a waveguide, regardless of the waveguide length, dispersion, or temporal duration.  \\
\section{Simulations of Higher-Order Dispersion}
\indent The first step in this effort is to focus on higher-order nonlinearities as they disturb fundamental soliton pulse propagation in an optical waveguide.  The model was set-up in order to determine the changes in Time Bandwidth Product (TBP) \cite{29} of fundamental solitons propagating with a range of $\mathrm{2^{nd}}$-order GVD coefficients, spanning from low-dispersion fiber waveguides to highly-dispersive PhCWG.  The study encompassed $\mathrm{3^{rd}}$, $\mathrm{4^{th}}$, $\mathrm{5^{th}}$, and $\mathrm{6^{th}}$-order GVD, in order to determine the maximum higher-order GVD term that the fundamental soliton can sustain itself over a given nonlinear length scale.  \\
\indent The simulation was built to analyze waveguides with $\mathrm{2^{nd}}$-order GVD coefficients that ranged from $\mathrm{10^{-5}}$ to 10 $\mathrm{ps^{2}/mm}$.  The simulation ran in 2 fs increments from 2 fs to 1 ps.  Each higher-order GVD coefficient was increased independently and incrementally from no higher-order dispersion to 0.1 $\mathrm{ps^{3}/mm}$, 0.05 $\mathrm{ps^{4}/mm}$, -0.3 $\mathrm{ps^{5}/mm}$, and -1 $\mathrm{ps^{6}/mm}$, as the maximum $\mathrm{3^{rd}}$, $\mathrm{4^{th}}$, $\mathrm{5^{th}}$, and $\mathrm{6^{th}}$-order GVD coefficient.  Equivalent waveguide lengths of 0.01 to 10 nonlinear length ratios were used in the study. Each simulation propagated an ideal, transform-limited, hyperbolic-secant-squared pulse function, with a calculated pulse energy determined from equation \ref{eq:MyEquToCite2} for a fundamental soliton. \\
\indent After the simulations, the data was analyzed to determine the maximum higher-order dispersion that could be tolerated for a fundamental soliton of a given temporal duration.  At each specified pulse duration, nonlinear length, and $\mathrm{2^{nd}}$-order GVD, the fundamental soliton was considered to be disturbed if the TBP changed by more than 10\%.  It was observed that the waveguides with highly-dispersive $\mathrm{2^{nd}}$-order GVD values were more tolerant to a given higher-order GVD.  This is expected, as a fundamental soliton within a highly-dispersive waveguide will have a greater input energy and thus greater SPM to balance out the increased group-velocity dispersion.  In addition, longer waveguide lengths were observed to be more sensitive to higher-order GVD; this is expected as the dispersion is proportional to the propagation length.  \\
\indent After the simulations were conducted, an effort was made to determine if a mathematical function could be found of the maximum higher-order dispersion as a function of the temporal width, nonlinear length, and $\mathrm{2^{nd}}$-order GVD.  This is a straightforward linear algebra problem utilizing least squares \cite{24}: \\
\begin{eqnarray}
  {\hat{A}}{\bar{x}}={\bar{b}},\\
  {{\hat{A}}^{T}}{\hat{A}}{\bar{x}}={{\hat{A}}^{T}}{\bar{b}},\\
  {{({{\hat{A}}^{T}}{\hat{A}})}^{-1}}({{\hat{A}}^{T}}{\hat{A}}){\bar{x}}={\bar{x}}={{({{\hat{A}}^{T}}{\hat{A}})}^{-1}}{{\hat{A}}^{T}}{\bar{b}},\
\end{eqnarray}
where ${\hat{A}}$ (this matrix is not to be confused with \emph{A(z,T)} or ${\bar{A}(z,\omega)}$, the pulse envelope function) represents the values of the individual functions to curve fit with the independent variables (temporal width, $\mathrm{2^{nd}}$-order GVD, and nonlinear length ratio), ${\hat{A}^{T}}$ is the transpose of ${\hat{A}}$, ${{\bar{x}}}$ represents the constant coefficients of these functions, and ${{\bar{b}}}$ represents the dependent data of the study (the maximum higher-order GVD).  The function found to most-closely represent the data is: 
\begin{equation}
  f(\tau,\beta_{2},L_{NL})={C_{1}}{\cdot}{{f_{1}}(\tau,\beta_{2},L_{NL})}+{C_{2}}{\cdot}{{f_{2}}(\tau,\beta_{2},L_{NL})}+{C_{3}},\\
\end{equation}
\begin{equation}
\label{eq:MyEquToCite}
  f(\tau,\beta_{2},L_{NL})={C_{1}}{\cdot}{\tau}{\cdot}{exp(-L_{NL})}+{C_{2}}{\cdot}{\beta_{2}}{\cdot}{exp(-L_{NL})}+{C_{3}},\\
\end{equation}
where $\tau$ (s) is the temporal duration, $\beta_2$ (s$\mathrm{^2}$/m) is the 2$\mathrm{^{nd}}$-order GVD coefficient, $L_{NL}$ is the ratio of the propagation length over the nonlinear length, calculated with equation \ref{eq:MyEquToCite3}, and $C_1$, $C_2$, and $C_3$ are the coefficients of the least squared method.   

\begin{table}[h!]
\begin{center}
\begin{tabular}{ | c | c | c | c | c | }
  \hline                        
  {} & {$\mathrm{3^{rd}}$-order} & {$\mathrm{4^{th}}$-order} & {$\mathrm{5^{th}}$-order} & {$\mathrm{6^{th}}$-order} \\
  \hline
  {$C_{1}$} & {$1.8770079{\cdot}10^{-22}$} & {$3.3950815{\cdot}10^{-35}$} & {$-2.3322422{\cdot}10^{-46}$} & {$-6.9359041{\cdot}10^{-58}$}  \\
  \hline
  {$C_{2}$} & {$-6.0439270{\cdot}10^{-15}$} & {$-4.0927729{\cdot}10^{-28}$} & {$2.1436889{\cdot}10^{-39}$} & {$7.5262895{\cdot}10^{-51}$} \\
  \hline
  {$C_{3}$} & {$1.2001781{\cdot}10^{-34}$} & {$1.5676273{\cdot}10^{-47}$} & {$-8.0129045{\cdot}10^{-59}$} & {$-2.4052190{\cdot}10^{-70}$} \\
  \hline  
  {Error} & {0.2421\%} & {0.0924\%} & {0.0845\%} & {0.1269\%}  \\
  \hline  
\end{tabular}
\caption{Coefficients for Equation \ref{eq:MyEquToCite}}
\end{center}
\end{table}
\indent By solving the least-squares equation, one can find the coefficients of the chosen equation, which are listed in Table 1.  The average error between equation \ref{eq:MyEquToCite} and the simulated maximum higher-order GVD tolerances was found to be 0.1365\%, thus validating the equation for the range of parameters studied.  By following equation \ref{eq:MyEquToCite}, one can reasonably predict the maximum higher-order dispersion possible for a fundamental soliton propagating in a waveguide with a given $\mathrm{2^{nd}}$-order GVD, temporal width, and nonlinear length ration.  \\

\begin{figure}[h]
\begin{minipage}[b]{0.45\linewidth}
\centering
\includegraphics[width=\textwidth]{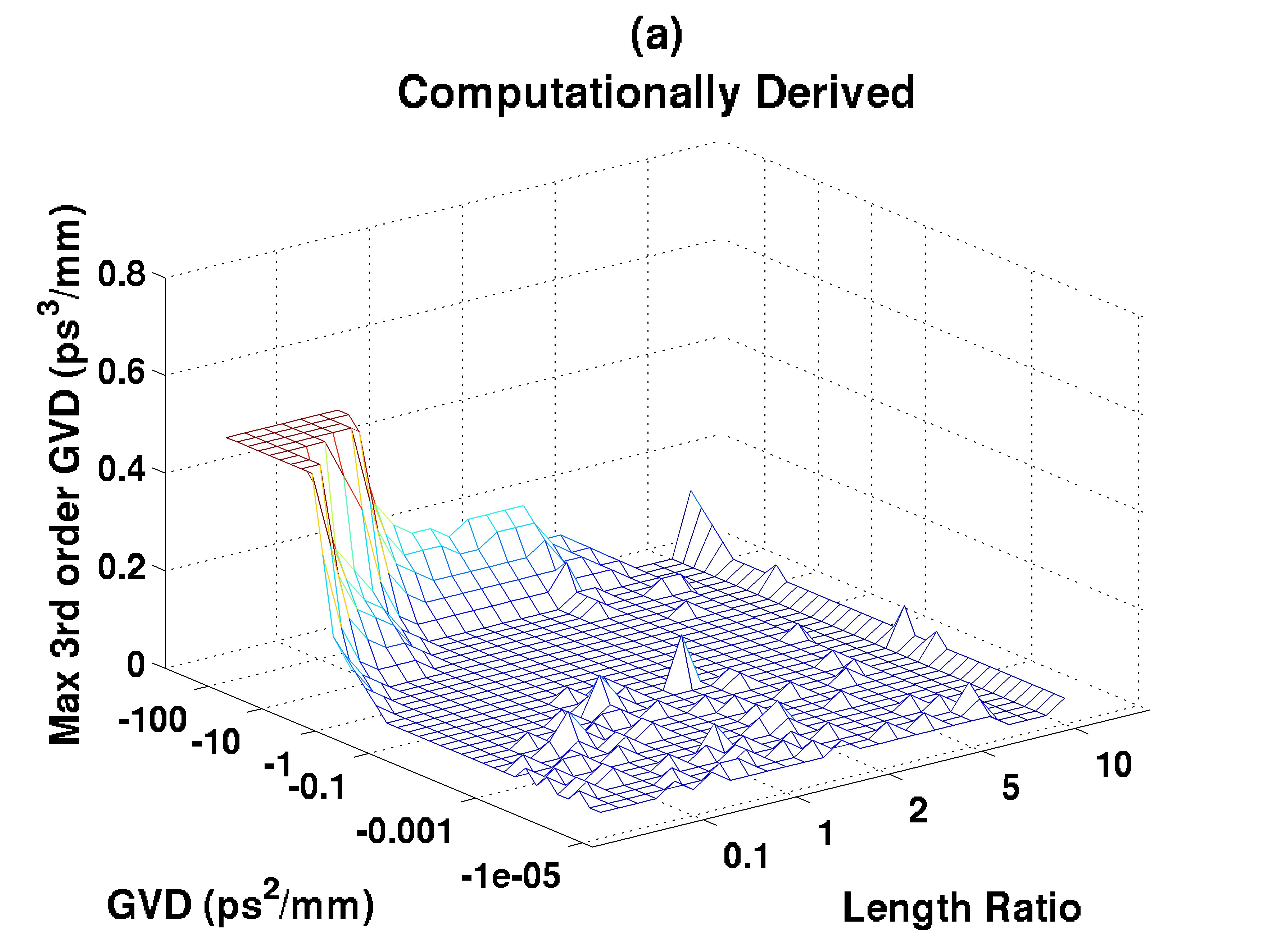}
\end{minipage}
\begin{minipage}[b]{0.45\linewidth}
\centering
\includegraphics[width=\textwidth]{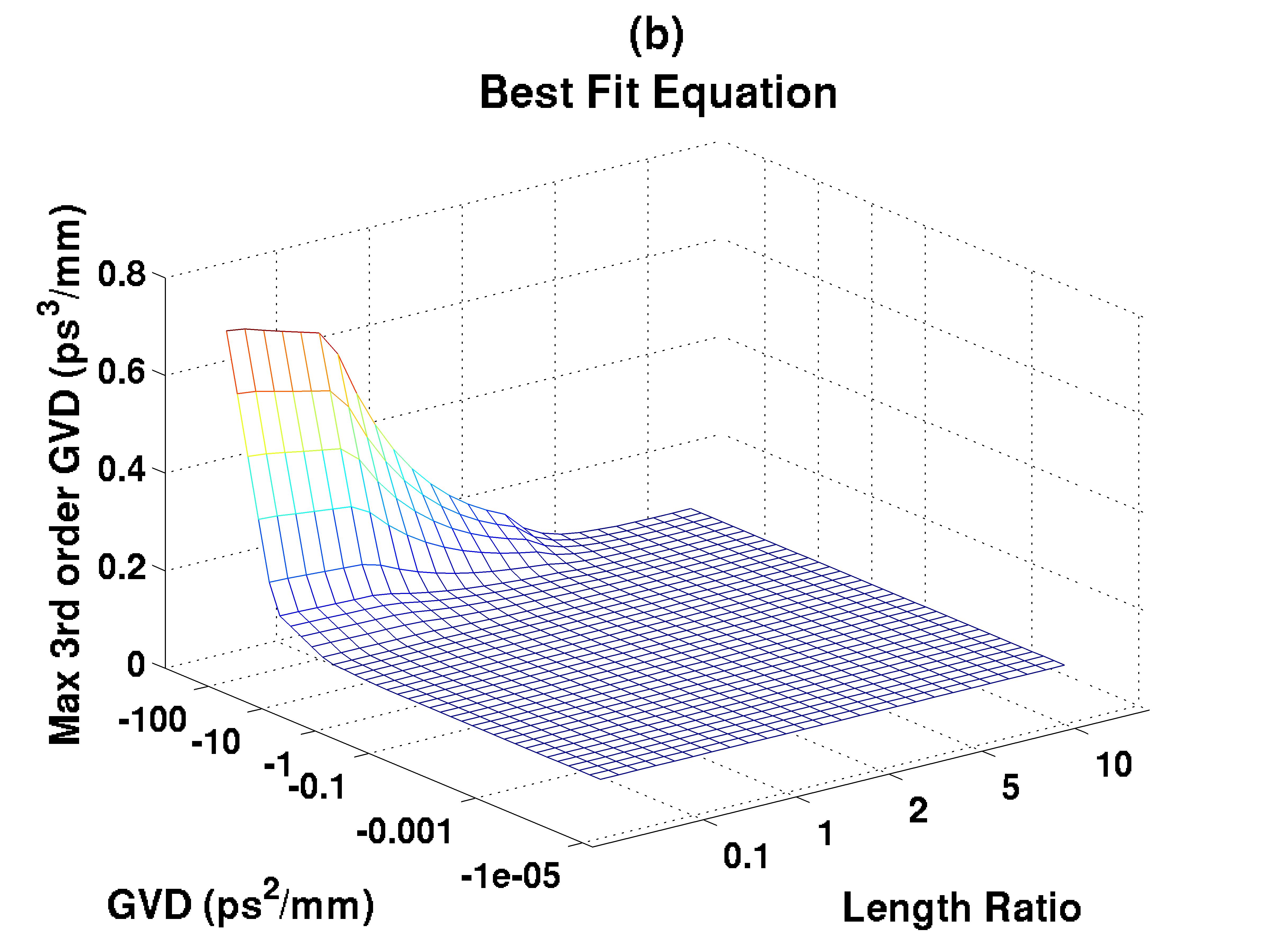}
\end{minipage}\\
\caption{Maximum $\mathrm{3^{rd}}$-order GVD coefficients for a 2 fs pulse, as a function of nonlinear length and $\mathrm{2^{nd}}$-order GVD, obtained through (a) direct NLSE simulations and (b) equation \ref{eq:MyEquToCite}. }
\label{fig:fig2}
\end{figure}

\section{Photonic Crystal Waveguide Analysis}

\indent In practice, there will always be some higher-order dispersion, and in a PhCWG this can be significant enough to measure experimentally.  In the analysis of the GaInP chip, the third order dispersion (TOD) was measured up to 0.1 $\mathrm{ps^{3}/mm}$ \cite{3}.  The next step in this effort was to run numerous NLSE simulations of the fundamental soliton, for various temporal durations and TOD, and to determine whether or not the self-sustaining characteristics can exist in the presence of TOD.  \\
\indent PhCWG's are, by definition, a periodic series of changing dielectric materials \cite{9,10}, such as a silicon slab with a repeating pattern of small air holes \cite{11}.  By engineering the periodic structure, one can engineer a $\emph{photonic band-gap}$, which effectively works as a frequency-selective mirror, which can be used to guide and contain an optical pulse.  At wavelengths near the edge of this photonic band-gap, the group delay increases dramatically as a function of wavelength \cite{8}, resulting in extraordinary geometric dispersion.  In addition, this increased group delay, also known as $\emph{slow-light}$, enhances the nonlinear Kerr effect.  The combination of this greatly enhanced dispersion and Kerr enables soliton evolution within extremely short waveguide lengths \cite{3}.  \\
\indent The model was first set at ten nonlinear lengths.  As expected, when there was zero TOD, there was no change in output TBP or output phase standard deviation, regardless of the input pulse duration; the TBP of the output was always equal to that of the input.  When the dispersion was increased to 0.1 $\mathrm{ps^{3}/mm}$, there was a noticeable change in the temporal duration for input temporal durations less than 100 fs.  The TBP would vary from half the input TBP, which is evidence of higher-order soliton compression as a result of the increased dispersion; and up to over twice the TBP, which is evidence of even greater pulse broadening and distortion of the pulse shape.  \\
\begin{figure}[h]
\centering
\begin{minipage}[b]{0.45\linewidth}
\centering
\includegraphics[width=\textwidth]{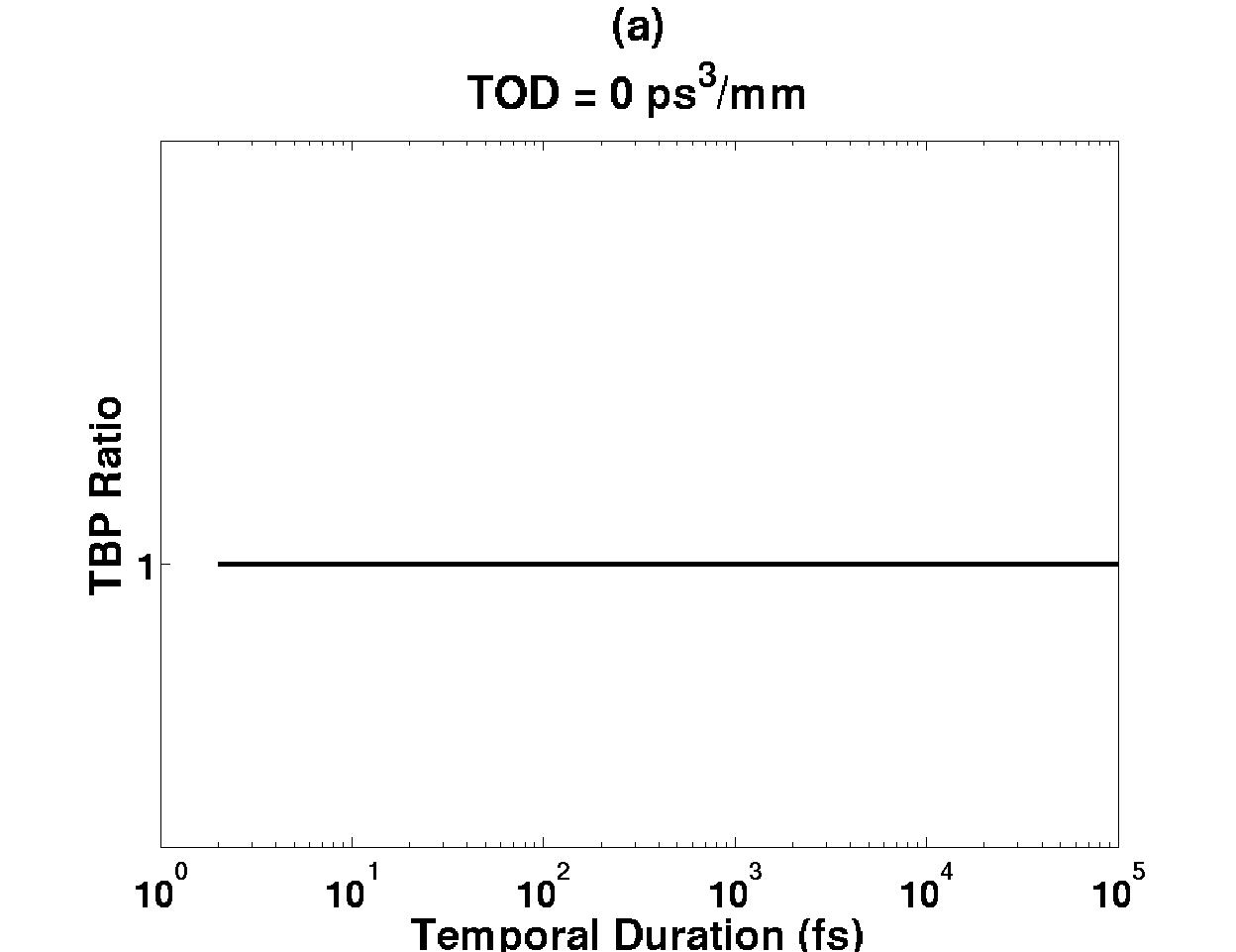}
\end{minipage}
\begin{minipage}[b]{0.45\linewidth}
\centering
\includegraphics[width=\textwidth]{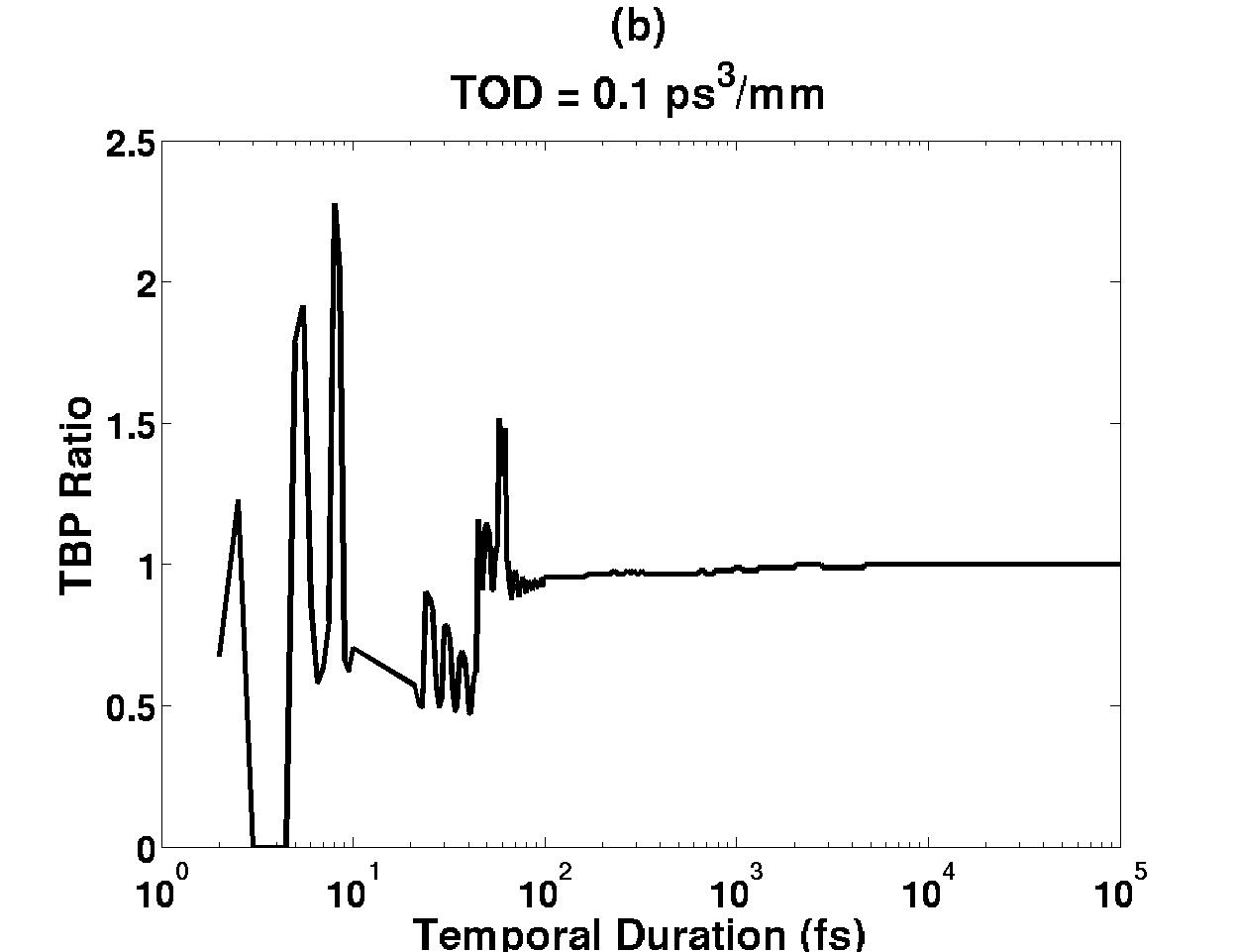}
\end{minipage}
\caption{The output TBP over the input TBP for an optical pulse propagating 10 nonlinear lengths, with an input pulse energy equal to that of the fundamental soliton, and with temporal durations ranging from 1 fs to 100 ps.  The TOD has a value of (a) 0 $\mathrm{ps^{3}/mm}$, and (b) 0.1 $\mathrm{ps^{3}/mm}$. }
\label{fig:fig3}
\end{figure}
\indent The analysis was then repeated for different ratio's of nonlinear lengths, and at numerous different TOD, ranging from 0 to 0.2 $\mathrm{ps^{3}/mm}$.  Again, the standard selected in order to consider a soliton pulse sustained within the waveguide was an output TBP deviation of less than 10\% of the input TBP.  It was observed that the minimum time would increase from no minimum time (as is the case of no TOD) almost linearly proportional to the TOD.  \\
\begin{figure}[h]
\centering
\begin{minipage}[b]{0.45\linewidth}
\centering
\includegraphics[width=\textwidth]{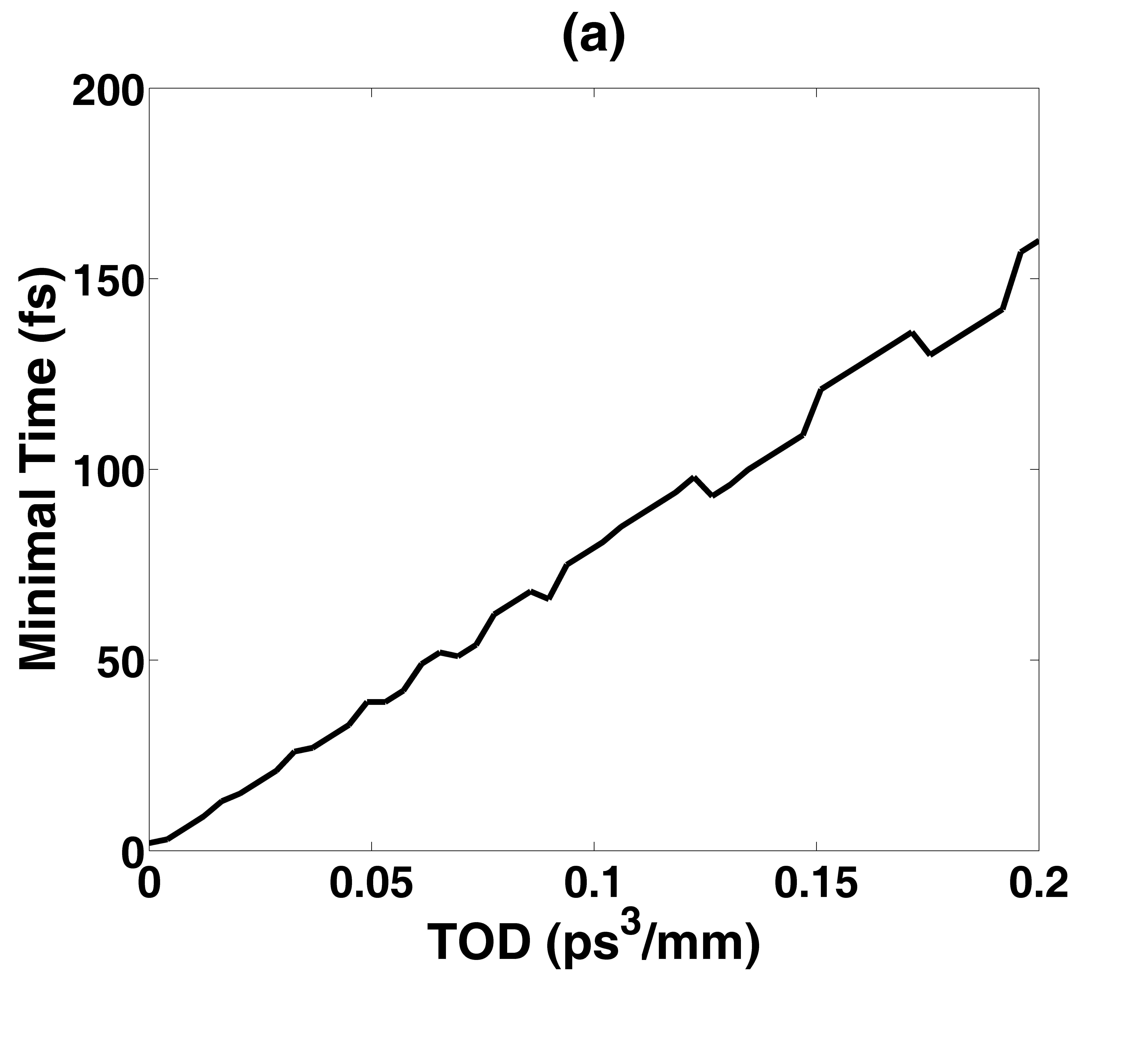}
\end{minipage}
\begin{minipage}[b]{0.45\linewidth}
\centering
\includegraphics[width=\textwidth]{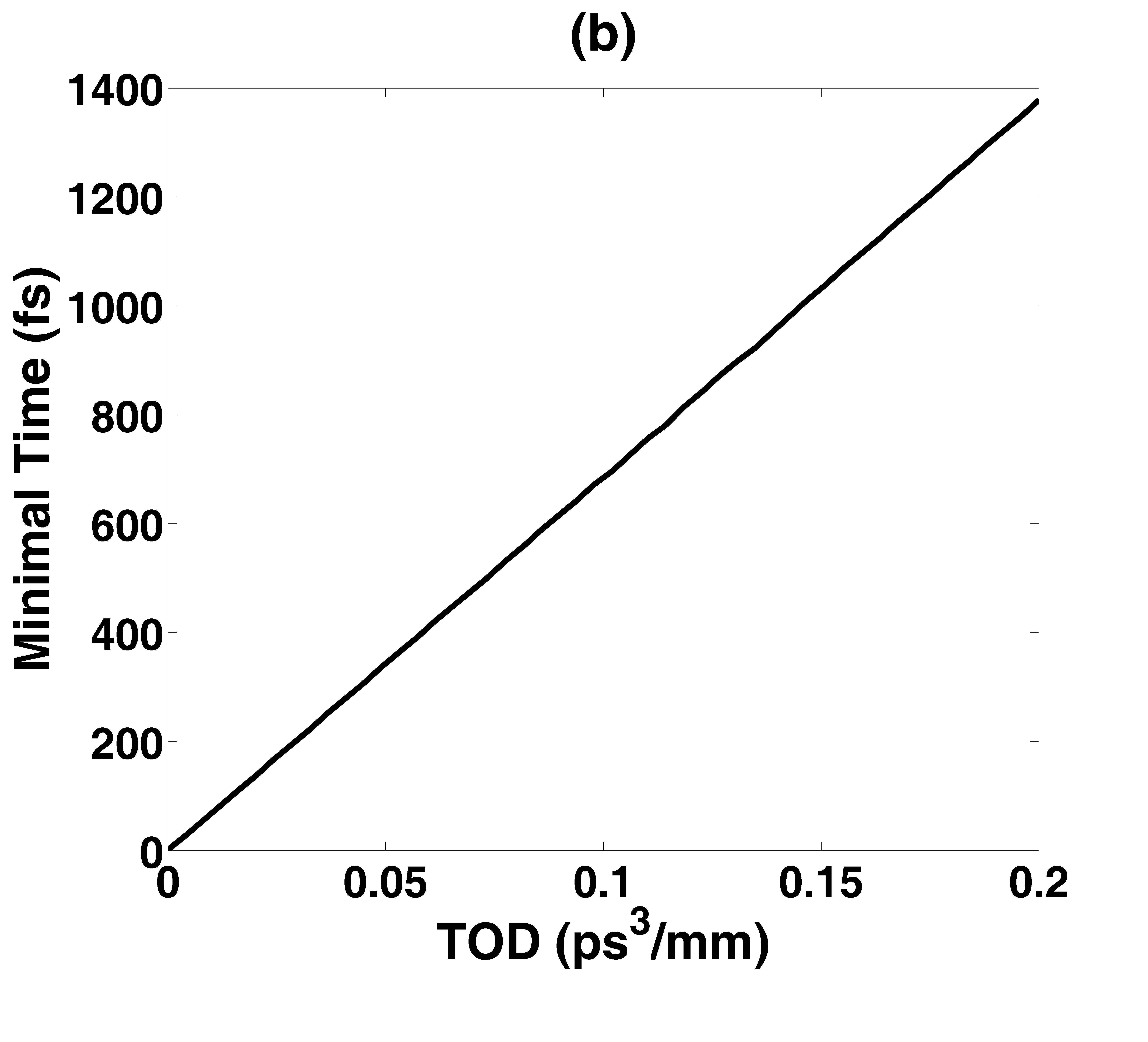}
\end{minipage}
\caption{Data plots of soliton disturbance as a function of TOD for 10 nonlinear lengths, both for (a) TBP and (b) phase. }
\label{fig:fig4}
\end{figure}
\indent In the next step, a similar analysis was conducted relating to the disturbance of the phase of the output of the PhCWG.  Throughout the study, a transform-limited pulse with no chirp was consistently used, where a phase of 0 radian was applied to the entire pulse.  On the output, it is expected that the phase would be shifted due to the SPM, and so the standard-deviation of the output pulse was determined.  In the absence of TOD, there was no change in the standard-deviation of the output pulse, regardless of the input temporal duration; this is expected as the pulse has a consistent shape as a fundamental soliton.  The standard selected for disturbance in the phase was a deviation of 0.1 radians of the standard-deviation of the output phase, when compared to the output phase at the longest 100 ps temporal duration. As the phase should remain constant for a fundamental soliton, deviation in the temporal phase (determined by the standard deviation) would be evidence of soliton disturbance. By increasing the TOD, it was observed that the minimum time would also increase linearly proportional with the TOD; the same as the minimum time for the soliton pulse stability.  For a single nonlinear length, the minimum time was a maximum of 70 fs (for a TOD of 0.2 $\mathrm{ps^{3}/mm)}$; whereas for ten nonlinear lengths, the minimum time for a stable phase was 1.3 ps (for the same TOD).  It is observed from the numerical study that the soliton phase stability is much more sensitive to TOD than the soliton intensity shape.  In practice, however, most PhCWG are less than a nonlinear length, so it is expected that, for pulses longer than 100 fs, there will be little phase disturbance as a result of TOD in practical applications of soliton pulse propagation within PhCWG.  \\

\section{Conclusion}
\indent In conclusion, a numerical study was conducted to theoretically understand what will happen to an ideal ultrafast fundamental soliton pulse propagating in a waveguide subjected to higher-order GVD.  Simulations were conducted for fundamental solitons subjected to a large range of $\mathrm{3^{rd}}$, $\mathrm{4^{th}}$, $\mathrm{5^{th}}$, and $\mathrm{6^{th}}$-order GVD, propagating with fs durations and $\mathrm{2^{nd}}$-order GVD ranging from low-dispersion fibers to highly dispersive PhCWG.  Throughout the study, the pulses were considered disturbed if the TBP changed by more than 10\%.  The study then turned to focus on practical applications of PhCWG, and it was determined that, when the waveguide has TOD typical to a modern PhCWG, the soliton will become disturbed at pulses shorter than 100 fs.  The phase of a fundamental soliton, which is subjected to self-focusing from SPM, was observed to become disturbed by the TOD at temporal durations ranging from 70 fs for a single nonlinear length, up to over a picosecond for ten nonlinear lengths; since most PhCWG are less than a full nonlinear length, TOD phase disturbance should not become an issue in practical applications for pulses under 100 fs.  In conclusion, the study demonstrated that fundamental soliton pulses longer than a hundred fs could propagate undistorted in a practical PhCWG while subjected to higher-order GVD.  By use of these PhCWG, one can accurately control and sustain these ultrafast, highly intense pulses over long distances, a phenomena that has a host of applications ranging from low-energy optical data transfer to high-energy pulsed lasers systems.  \\

\section{Acknowledgements}

\indent Sources of funding for this effort include Navy Air Systems Command (NAVAIR)-4.0T Chief Technology Officer Organization as an Independent Laboratory In-House Research (ILIR) Basic Research Project (Nonlinear Analysis of Ultrafast Pulses with Modeling and Simulation and Experimentation); a National Science Foundation (NSF) grant (Ultrafast nonlinearities in chip-scale photonic crystals, Award \#1102257), the Intelligence Community Postdoctoral Fellowship, the National Science Foundation of China (NSFC) award \#61070040 and \#60907003, and the Science Mathematics And Research for Transformation (SMART) fellowship.  The authors thanks Tingyi Gu, Michael Weinstein, William Torruellas, James McMillan, Jiali Liao, Ying Li, and Chad Husko for fruitful discussions.   \\

\end{document}